\documentclass[conference]{IEEEtran}

\IEEEoverridecommandlockouts

\usepackage{cite}
\usepackage{amsmath,amssymb,amsfonts}
\usepackage{algorithmic}
\usepackage{graphicx}
\usepackage{textcomp}
\usepackage{xcolor}
\usepackage{bm}

\usepackage{hyperref}
\hypersetup{
    colorlinks=true,
    citecolor=blue,
}
\usepackage{colortbl}

\usepackage{cite}
\usepackage{url}
\usepackage{amsfonts}
\usepackage{multirow}
\usepackage{makecell}
\usepackage{booktabs} 
\usepackage{subcaption}
\usepackage{xspace}
\usepackage{array}

\usepackage{pifont}

\DeclareMathOperator*{\argmin}{arg\,min}
\newcommand{\bx}{\bm{x}}
\newcommand{\xo}{\bx_0}
\newcommand{\tildex}{\tilde{\bx}}
\newcommand{\xsp}{\tildex_{\text{sp}}}
\newcommand{\xqp}{\tildex_{\text{qp}}}
\newcommand{\lsp}{\lambda_{\text{sp}}}
\newcommand{\lqp}{\lambda_{\text{qp}}}
\newcommand{\dqp}{\Delta_{\text{qp}}}
\newcommand{\wav}{\texttt{waveform}\xspace}
\newcommand{\mhfg}{\texttt{mel-spec + HiFi-GAN}\xspace}
\newcommand{\encodec}{\texttt{EnCodec latent}\xspace}

\ifCLASSOPTIONpeerreview
  \newcommand{\demourl}{\url{https://anonymous.4open.science/w/attack-utmos-demo-01E4/}}
\else
  \newcommand{\demourl}{\url{https://unilight.github.io/attack-utmos-demo/}}
\fi

\begin{document}

\title{Attacking UTMOS: Probing the Robustness of a Speech Quality Assessment Model}

\author{
    \IEEEauthorblockN{
        \textit{
            Wen-Chin Huang, Tomoki Toda
        }
    }
    \IEEEauthorblockN{
        Nagoya University, Japan
    }
}

\ifCLASSOPTIONpeerreview
    \IEEEpeerreviewmaketitle
\else
    \maketitle
\fi

\begin{abstract}
UTMOS has become one of the most commonly used deep neural network-based speech quality assessment (SQA) metrics in speech processing research. In this paper, we attack UTMOS to probe its robustness. Starting from high-quality speech samples, we optimize the input in two directions: a score-preserving attack, which degrades perceived quality while maintaining the predicted score, and a quality-preserving attack, which lowers the predicted score while maintaining perceived quality. We consider three input spaces: raw waveform, mel spectrogram with a HiFi-GAN vocoder, and the latent space of EnCodec, a neural audio codec. Experimental results show that score-preserving attacks are effective against UTMOS. Although perfect quality-preserving attacks are more difficult, optimization in the EnCodec latent space provides the best chance of success. These results reveal failure modes of UTMOS and highlight the importance of robustness analysis for DNN-based SQA metrics.
\end{abstract}

\begin{IEEEkeywords}
speech quality assessment, UTMOS, robustness, adversarial example, adversarial attack
\end{IEEEkeywords}

\section{Introduction}
\label{sec:intro}

Speech quality assessment (SQA) refers to the evaluation of speech quality \cite{speech-evaluation-review-2011, speech-evaluation-review-2024}.
In this work, we are particularly interested in non-intrusive, deep neural network (DNN)-based methods trained with $\langle \text{speech}, \text{quality label}\rangle$ pairs, where the quality labels are collected through mean opinion score (MOS)-style listening tests \cite{mosnet, nisqa, dnsmos, ssl-mos}\footnote{In the remainder of the paper, unless specified, the term ``SQA model'' refers to this type of method.}.
One of the most popular SQA models is UTMOS \cite{utmos}, the winning system of the VoiceMOS Challenge 2022 \cite{voicemos2022}. Trained on the BVCC dataset \cite{bvcc}, which contains synthetic speech samples from 187 text-to-speech (TTS) and voice conversion (VC) systems, UTMOS was shown to have a system-level Spearman's rank correlation coefficient larger than 0.9 with human ratings \cite{voicemos2022}.

The strong performance of DNN-based SQA models has led to their wide adoption in speech research. Their use can be broadly categorized into two directions. The first is to use them as evaluation metrics. While this use is most direct in speech synthesis research \cite{naturalspeech3}, these models have also been adopted in other speech processing tasks, including speech enhancement \cite{cmgan}, neural audio coding \cite{wavtokenizer}, and spoken dialogue systems \cite{LLaMA-Omni}. The second, more recent direction is to use SQA models to guide the training of speech generation models, by using them as loss functions for TTS \cite{perceptual-tts} or reward functions to perform preference alignment \cite{preference-alignment-llm-tts}, or as training objectives for other speech generation tasks, such as speech enhancement \cite{se-multimetric}.

The increasing use of SQA models has raised the need to carefully examine their robustness. Due to their data-driven nature, models such as UTMOS can suffer from robustness issues, with out-of-domain (OOD) generalization being one of the most widely studied aspects. Previous studies have repeatedly shown that SQA models can generalize poorly to speech samples from different domains, such as noisy speech or singing voices; different languages, such as Chinese or French; newer and higher-quality speech synthesis systems; or even different listening-test settings \cite{voicemos2023, mos-bench}. In particular, such domain shifts can reduce their ability to correctly rank the performance of speech synthesis systems. These findings have led researchers to call for more careful use of SQA models to evaluate speech generation systems \cite{cooper2025good}.

\begin{figure}
    \centering
    
    \includegraphics[width=0.75\linewidth]{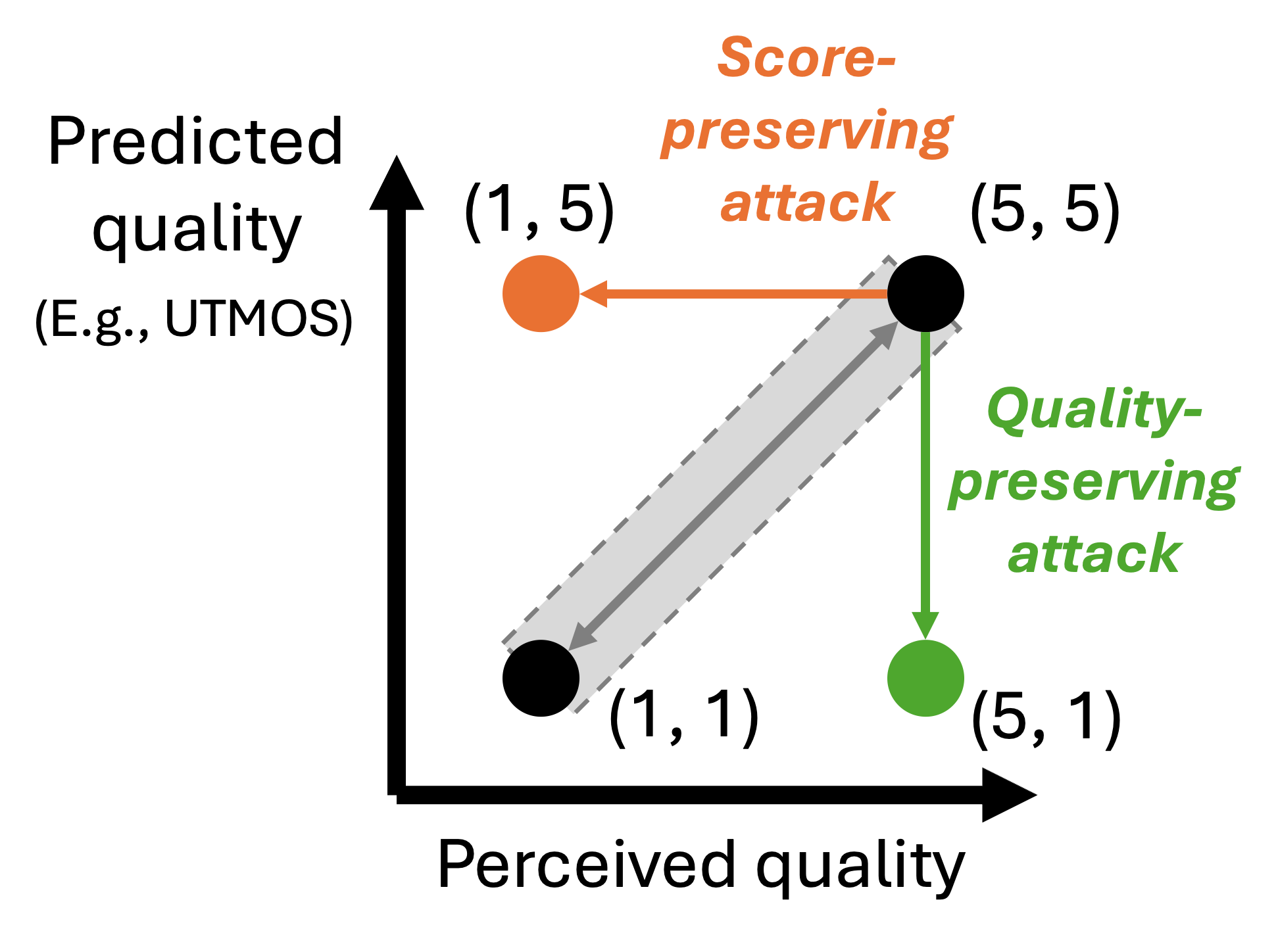}
    \caption{
        Illustration of the desired behavior zone of an ideal speech quality aseessment model (grey arrow), and the proposed score-preserving attack (orange arrow) and the quality-preserving attack (green arrow) directions.
    }

    \label{fig:illustration}
    \vspace{-5pt}
\end{figure}

    

The line of research discussed above mainly identifies \textit{naturally occurring} examples that expose robustness issues in SQA models. In this work, we take a complementary approach: we actively \textit{engineer samples to attack} SQA models, focusing specifically on UTMOS, and construct examples that challenge the model \textit{by design}. This idea is closely related to adversarial examples generated through input optimization \cite{first-ae}. Fig.~\ref{fig:illustration} illustrates the basic concept. An ideal SQA model should follow the gray arrow: speech with low perceived quality should receive a low predicted quality score, and vice versa. Based on this perspective, we propose two attacks against UTMOS. Starting from a high-quality speech sample, the first attack minimizes the perceived quality while maintaining the predicted quality score. We refer to this as a \textbf{score-preserving attack}. The second attack minimizes the predicted quality score while preserving perceived quality, which we refer to as a \textbf{quality-preserving attack}. Neither attack should succeed against an ideal and robust SQA model.

In the experiments, we instantiate the proposed attacks in three optimization spaces: the waveform space, the latent space of a neural audio codec, EnCodec \cite{encodec}, and the mel-spectrogram space coupled with a neural vocoder, HiFi-GAN \cite{hifigan}.
The results show that UTMOS is vulnerable to score-preserving attacks, whereas quality-preserving attacks are substantially more difficult, with the EnCodec latent space providing the most promising results.
These findings suggest that the robustness issues of DNN-based SQA models are not limited to naturally occurring OOD samples, but can also be exposed through adversarially constructed examples. Audio samples are available online\footnote{\demourl}.

\section{Related works}

\subsection{White-box gradient-based adversarial attacks}
\label{ssec:related-white-box}

The phenomenon of adversarial vulnerability was first discovered in \cite{first-ae}, where it was demonstrated that applying imperceptible perturbations to inputs could cause DNNs to fail.
Early works heavily favored \textit{constrained attacks}, which treat the perturbation magnitude as a strict, hard boundary while maximizing the model loss \cite{fgsm, bim, deepfool, jsma, pgd}.
These methods forcibly clip the perturbation back into a predefined boundary at every optimization step.
Later, it was found that \textit{penalized attacks} could achieve higher success rates and tighter distortion bounds. The representative work is the C\&W attack \cite{cnw_attack}, which minimizes the perturbation while incorporating the adversarial objective as a soft loss penalty. These two goals are dynamically balanced via a weighting hyperparameter.
These works laid the foundations and mathematical groundwork for manipulating continuous signals (mostly image pixels), and researchers then adapted these concepts for various speech applications.

\subsection{Adversarial attacks in speech processing}
\label{ssec:related-ae-speech}



While speech processing has seen various forms of attacks, including attacks constructed to be perceivable by voice interfaces but inaudible to humans \cite{2016-hidden,2017-dophinattack,abdullah2019practical} and spoofing attacks against automatic speaker verification systems \cite{asvspoof2019, asvspoof2019-database}, the line of work most closely related to this paper is model-level adversarial examples. In these attacks, the input speech is optimized to change the prediction of a target speech model. Early studies generated targeted adversarial examples that forced automatic speech recognition (ASR) systems to output desired transcriptions, with later work improving perceptual imperceptibility or robustness to physical playback \cite{2018-ae-asr, 2019-ae-asr, ijcai2019p741}. Similar optimization-based frameworks have also been applied to speaker verification and recognition systems \cite{2018-ae-asv,2021-ae-asv}.

Attacks on speech quality assessment models differ from those on many other speech processing models and are non-trivial in an important way. 
Attacks on speech or speaker recognition models add small acoustic perturbations to change model predictions while, ideally, leaving high-level attributes such as phonetic content or speaker identity unchanged.
In contrast, \textbf{speech quality itself is not independent of low-level acoustic properties}. It can be affected by a broad spectrum of aspects: background noise, reverberation, intelligibility, and prosody. Therefore, directly perturbing the waveform can easily change the perceived quality, making the design of attacks on SQA models more difficult. This motivates us to consider not only waveform-space optimization, but also alternative attack spaces that may better control the trade-off between predicted quality and perceived quality.

\section{Methodology}

\subsection{Formulation}
\label{ssec:formulation}

Our attack starts with a high-quality\footnote{A low-quality speech sample may have multiple realizations, including background noise, transmission error, or artifacts from a TTS system. Investigating attacks that start with low-quality samples can thus be complicated, and we leave it for future work.} speech sample $\xo\in\mathcal{X}$ where $\mathcal{X}$ denotes the space of all possible speech realizations.
We denote the \textbf{perceived quality} (e.g., naturalness score from a five-point scale MOS test) of a speech sample as $f: \mathcal{X} \rightarrow [1, 5]$. Similarly, we denote the \textbf{predicted quality score} yielded by a SQA model as $g: \mathcal{X} \rightarrow [1, 5]$. Since $\xo$ is a clean, natural sample, we assume $f(\xo)=g(\xo)=5$. In this work, we specifically choose to attack the UTMOS model \cite{utmos}, which takes a speech waveform as input and outputs a score ranging from one to five.

We take inspiration from \cite{cnw_attack} and formulate the problem of attacking the SQA model as an untargeted optimization problem that can be solved iteratively. As illustrated in Fig.~\ref{fig:illustration}, we consider the following two types of attack directions:
\begin{itemize}
    \item A \textbf{score-preserving attack} aims to find an example $\tildex$ that degrades the actual, perceived quality as much as possible while preserving the predicted quality score. In our setting, this attack tries to find a speech sample $\xsp$ that maximizes the difference between $f(\xsp)$ and $f(\xo)$ while minimizing the difference between $g(\xsp)$ and $g(\xo)$. Formally,
        \begin{equation}
            \label{eq:score-preserving-attack}
            \xsp=\argmin_{\bx} - \left\lVert f(\bx) - f(\xo) \right\rVert _2  + \lsp \cdot \left\lVert g(\bx) - g(\xo) \right\rVert _2.
        \end{equation}
    Here, the first term tries to maximize the perceived quality degradation, and the second term tries to preserve the predicted score. The second term could be seen as a \textit{penalty} \cite{cnw_attack}, and $\lsp$ serves as a hyperparameter that controls the strength of such a penalty. 
    \item A \textbf{quality-preserving attack} aims to find an example $\tildex$ that decreases the predicted quality score as much as possible while preserving the actual perceived quality. In other words, this attack tries to find a speech sample $\xqp$ that maximizes the difference between $g(\xqp)$ and $g(\xo)$ while minimizing the difference between $f(\xqp)$ and $f(\xo)$. Formally,
        \begin{equation}
            \label{eq:quality-preserving-attack}
            \xqp=\argmin_{\bx} - \left\lVert g(\bx) - g(\xo) \right\rVert _2 + \lqp \cdot \left\lVert f(\bx) - f(\xo) \right\rVert _2.
        \end{equation}
    Similar to above, $\lqp$ is applied to the penalty term, the quality-preserving criterion.
\end{itemize}
The hyperparameters $\lsp, \lqp$ are expected to control the trade-off between the attack effectiveness and the penalty. Taking the score-preserving attack as an example, if $\lsp$ is set too large, the optimization will not progress because the penalty term prioritizes preserving the predicted quality score too heavily. On the other hand, if $\lsp$ is set too small, the penalty term will have no effect, and the optimization may simply degrade the perceived quality indefinitely without preserving the predicted quality. We examine this trade-off in Sec.~\ref{sec:exp}

In both attack directions, we need to be able to calculate $\left\lVert f(\bx) - f(\xo) \right\rVert _2$ during every optimization iteration, but it would be impractical to ask human to make the judgement in every single iteration. Following the early adversarial examples literature, in practice we use the $L_2$ norm in the attack space, such that
\begin{equation}
    \left\lVert f(\bx) - f(\xo) \right\rVert _2 = \left\lVert \bx-\xo \right\rVert _2.
\end{equation}

Note that $\mathcal{X}$ denotes the speech space in which the optimization, or attack, is performed and is not limited to the waveform space. We explain our choice of attack space in the next subsection.

\begin{figure}
    \centering
    
    \includegraphics[width=\linewidth]{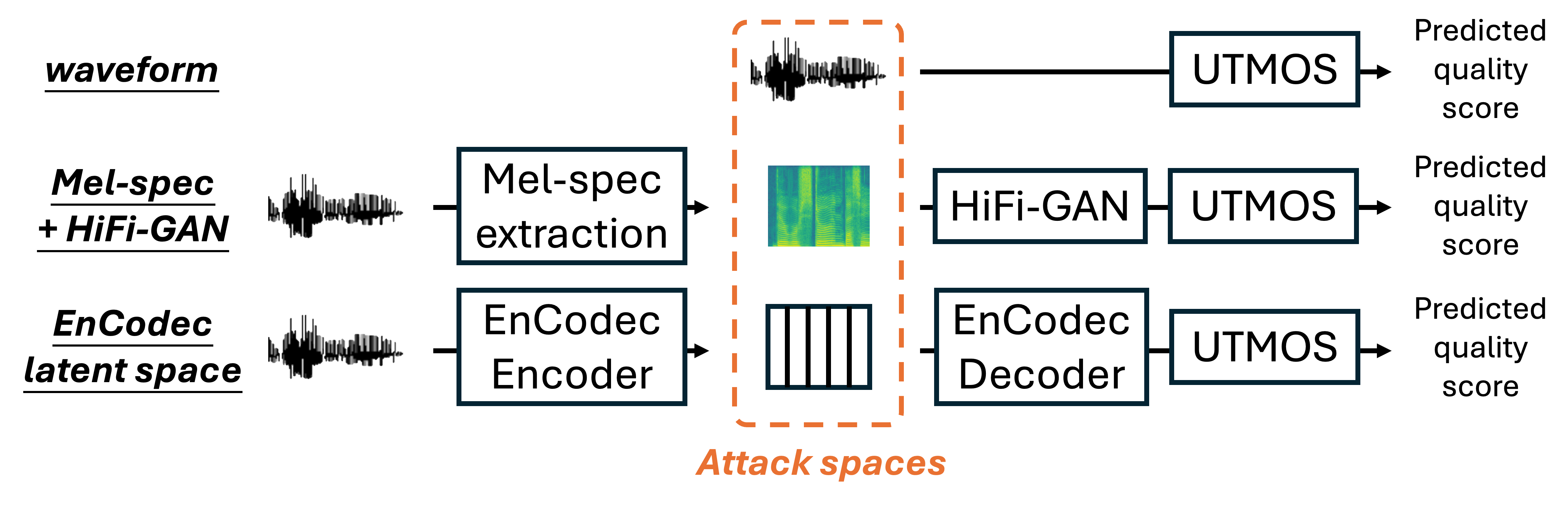}
    \caption{Illustration of the three attack spaces in this work.}

    \label{fig:attack-spaces}
\end{figure}

\subsection{Attack space}

In the adversarial examples literature where image is frequently used as a representative object, optimization typically works on the pixel space, and it was believed (and empirically shown) that two images with a sufficiently small $L_2$ distance in the pixel space are perceptually identical.
While a naive analogy would be to perform optimization in the waveform space, in this work we experiment with three input spaces to perform attack on, as shown in Fig.~\ref{fig:attack-spaces}.

\subsubsection{Waveform}

As explained above, the most straightforward idea is to simply input the speech waveform to the UTMOS model, and calculate the gradient with respect to the optimization objectives in either Eq.~\ref{eq:score-preserving-attack} or Eq.~\ref{eq:quality-preserving-attack} to perform input optimization on the speech waveform space. In the remainder of this paper, we denote this space as \wav. The problem with operating on the waveform is that because we are interested in the quality of the speech sample, even a small $L_2$ distance induced by a negligible noise on the waveform space can cause a strong quality difference.  

\subsubsection{Mel spectrogram with HiFi-GAN}

Our next choice is to operate on the space of mel spectrogram, which is a commonly used time-frequency representation that is easier to work on because of the lower temporal resolution and clearer acoustic structure. Since UTMOS takes waveform as input, we need to convert from mel spectrogram to waveform. In this work, we adopt the HiFi-GAN model \cite{hifigan}. In the remainder of this paper, we denote this space as \mhfg. During each optimization iteration, the HiFi-GAN model converts the mel spectrogram to waveform, which is further input to the UTMOS model to obtain the predicted quality score, and the gradient is finally propagated back to update the mel spectrogram.

Similar to waveform, a small $L_2$ distance on the mel spectrogram domain may still cause a perceptible quality difference, because phase is not considered. However, the adoption of the HiFi-GAN model brings an unexpected effect: because models like HiFi-GAN are typically trained on clean speech, theoretically it should induce a smooth speech manifold on the input space: even if we make wild modifications to the mel spectrogram, it is possible that HiFi-GAN always maps it to clean speech.

\subsubsection{EnCodec latent space}

Our final choice is to operate in the latent space of a neural audio codec, EnCodec \cite{encodec}. In the remainder of this paper, we denote this space as \encodec. EnCodec is an encoder--decoder model with a quantized latent representation, trained end-to-end on a wide variety of audio data. During each optimization step, the EnCodec decoder maps the latent codes back to a waveform, which is then passed to UTMOS to compute the predicted quality score. The gradient is then backpropagated through the decoder to update the latent representation.

Compared with \mhfg, the \encodec{} space has a larger degree of freedom because it does not need to follow the structure of the mel spectrogram. In addition, because EnCodec is trained not only on clean speech but also on noisy speech, music, and general audio, it is expected to impose a weaker speech manifold than \mhfg{} while still providing a more structured optimization space than the raw waveform.

\textbf{Why \mhfg{} and \encodec?} We choose these two spaces not only because they are equipped with decoders that induce speech-related manifolds, but also because they are widely used in modern TTS systems: HiFi-GAN is commonly used as the vocoder in two-stage, mel-spectrogram-based TTS systems \cite{matcha-tts, xtts}, whereas EnCodec is a key component of large language model-based TTS systems \cite{vall-e, voicecraft}. Demonstrating the vulnerability of UTMOS in optimization spaces associated with these models is therefore practically relevant.





\section{Experimental results}
\label{sec:exp}

\subsection{Experimental settings}

We utilized the \texttt{test-clean} subset of the LibriSpeech corpus \cite{librispeech}. We first randomly selected 30 distinct speakers. Then, for each speaker, we randomly sampled one speech utterance that met two criteria: (1) audio duration between three and five seconds, and (2) UTMOS score between four and five. All speech samples are in 16 kHz.

We used a wrapper of the official UTMOS model checkpoint\footnote{\url{https://github.com/tarepan/SpeechMOS}}, and we used the official EnCodec model checkpoint on HuggingFace\footnote{\url{https://huggingface.co/facebook/encodec_24khz}}. In addition, we used an open-source HiFi-GAN model checkpoint\footnote{\url{https://huggingface.co/speechbrain/tts-hifigan-libritts-16kHz}}, which was trained on the LibriTTS corpus \cite{libritts}.

All runs were optimized with Adam \cite{adam} for 50 iterations. The learning rate was set to be $1 \times 10^{-2}$ for \wav and $5 \times 10^{-2}$ for \mhfg and \encodec.
In practice, in the beginning of the optimization process, $\tildex$ is initialized as $\xo$. As a reuslt, the optimization objectives and thus the generated gradient in Eq.~\ref{eq:score-preserving-attack} and Eq.~\ref{eq:quality-preserving-attack} will be zero.
To avoid this, we modify the penality terms by adding a small $\epsilon$ offset to the target score or the input-domain penalty. This does not define a new attack goal; it only avoids degenerate zero-gradient behavior at the first optimization step.
The magnitude of $\epsilon$ was set to be $1 \times 10^{-4}$.

Perceived quality can ultimately only be evaluated by human listeners.
However, it is impractical to conduct listening tests for every attack configuration and optimization run.
We therefore use PESQ \cite{pesq} as a proxy metric for perceived quality in the experiments.
We avoid using another DNN-based SQA model, such as NISQA or DNSMOS, as the reference metric, since such models may themselves exhibit adversarial fragility.
Importantly, \textbf{PESQ is not treated as a substitute for human perception, but only as an objective proxy for screening and visualization}.
For the remainder of this section, we will be using PESQ-UTMOS plots for analysis.
Representative samples identified from the PESQ-based analysis are later validated through a listening test.

\begin{figure}
    \centering
    
    \includegraphics[width=\linewidth]{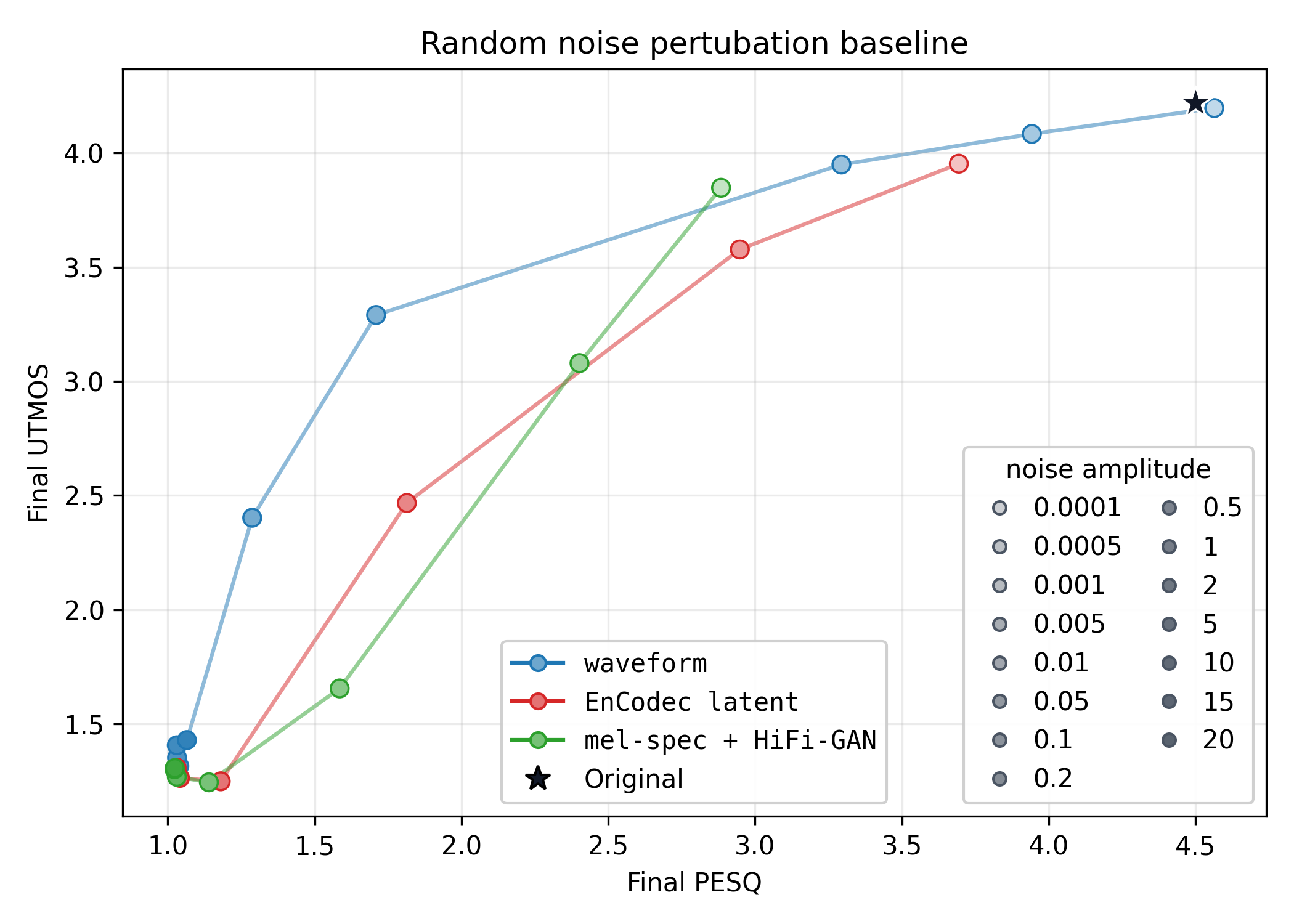}
    \caption{Random noise pertubation results. Random noise vectors with different amplitude were added to the three attack spaces, and the PESQ and UTMOS scores are calculated and plotted.}

    \label{fig:noise-baseline}
\end{figure}

\subsection{Random noise pertubation baseline results}
\label{ssec:noise-baseline}

One way to view the attacks is that we add perturbations to the input in a specific way to reach a certain goal. It therefore raises a fundamental question: how do the perceived and predicted quality change if we add random noise to the input? This should be viewed as the ``baseline''. To this end, we add Gaussian noises with different amplitude values to the three attack spaces. Figure~\ref{fig:noise-baseline} shows the results. It could be observed that as the noise amplitude increased, both PESQ and UTMOS scores dropped, which was in fact the desired behavior as depicted in Fig.~\ref{fig:illustration}. On the other hand, this result justifies our proposed attacks, as we aim to find samples that minimizes the perceived quality while preserving the predicted quality score, and vice versa.

\begin{figure}
    \centering
    
    \includegraphics[width=\linewidth]{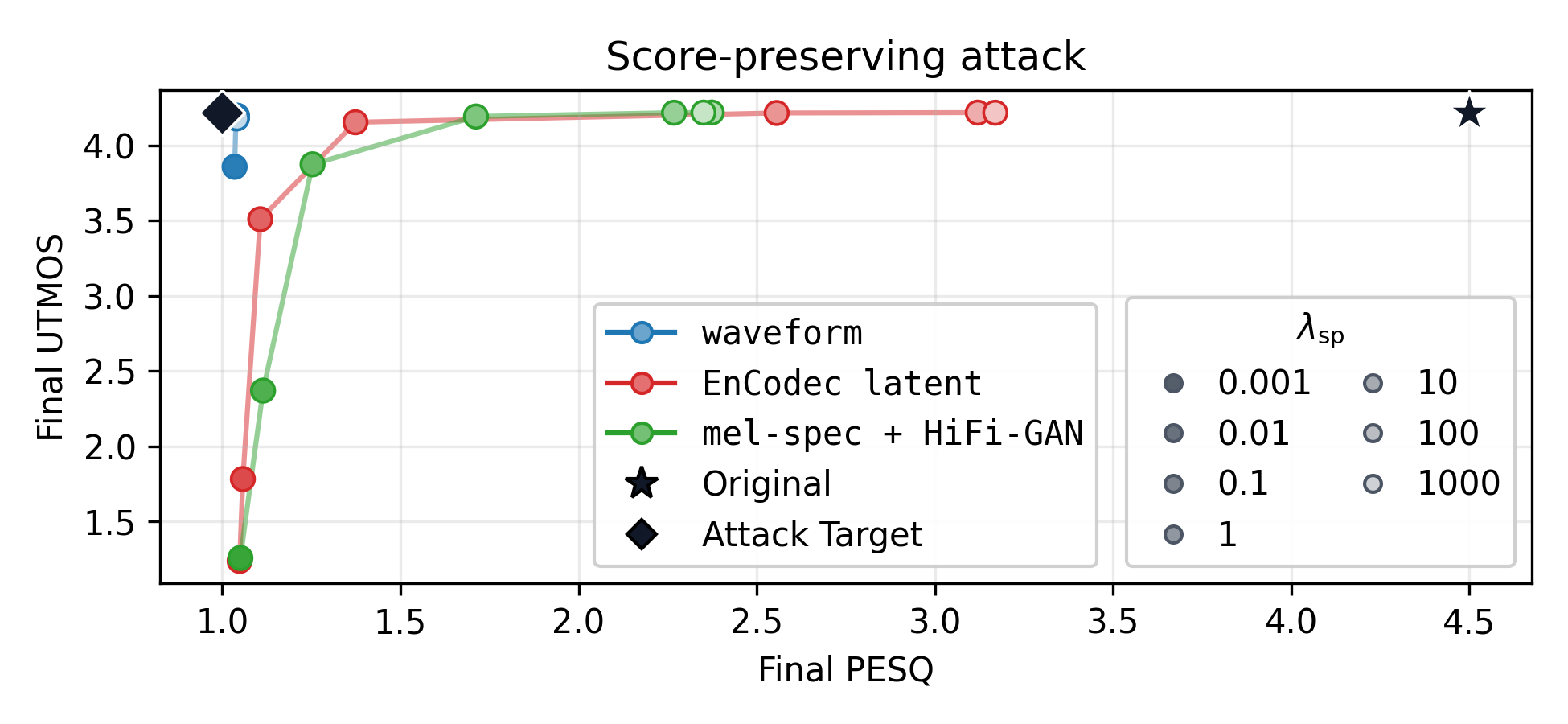}
    \caption{Score-preserving attack results.}

    \label{fig:score-preserving-results}
\end{figure}

\subsection{Score-preserving attack results}
\label{ssec:score-preserving-attack-results}

\subsubsection{Results from different attack spaces}

We now start with our first attack direction, the score-preserving attack. Fig.~\ref{fig:score-preserving-results} shows the results from different attack spaces and $\lsp$ values, which represents the stength of the score preservation penalty in Eq.~\ref{eq:score-preserving-attack}.
All three attack spaces produced samples located near the desired score-preserving region, i.e., low PESQ relative to the original sample, while maintaining a UTMOS score close to the attack target. This indicates that UTMOS can be insensitive to substantial signal degradation. Looking at individual spaces, for \wav, all $\lsp$ values gave similar PESQ and UTMOS results. On the other hand, we observed different outcomes with different $\lsp$ values in the \encodec and \mhfg space: as $\lsp$ decreases (i.e., the score-preservation penalty weakens) and PESQ score drops, the final UTMOS score decreases as well.

\begin{figure}[t]
  \centering
  \begin{subfigure}{\linewidth}
    \centering
    \includegraphics[width=0.85\linewidth]{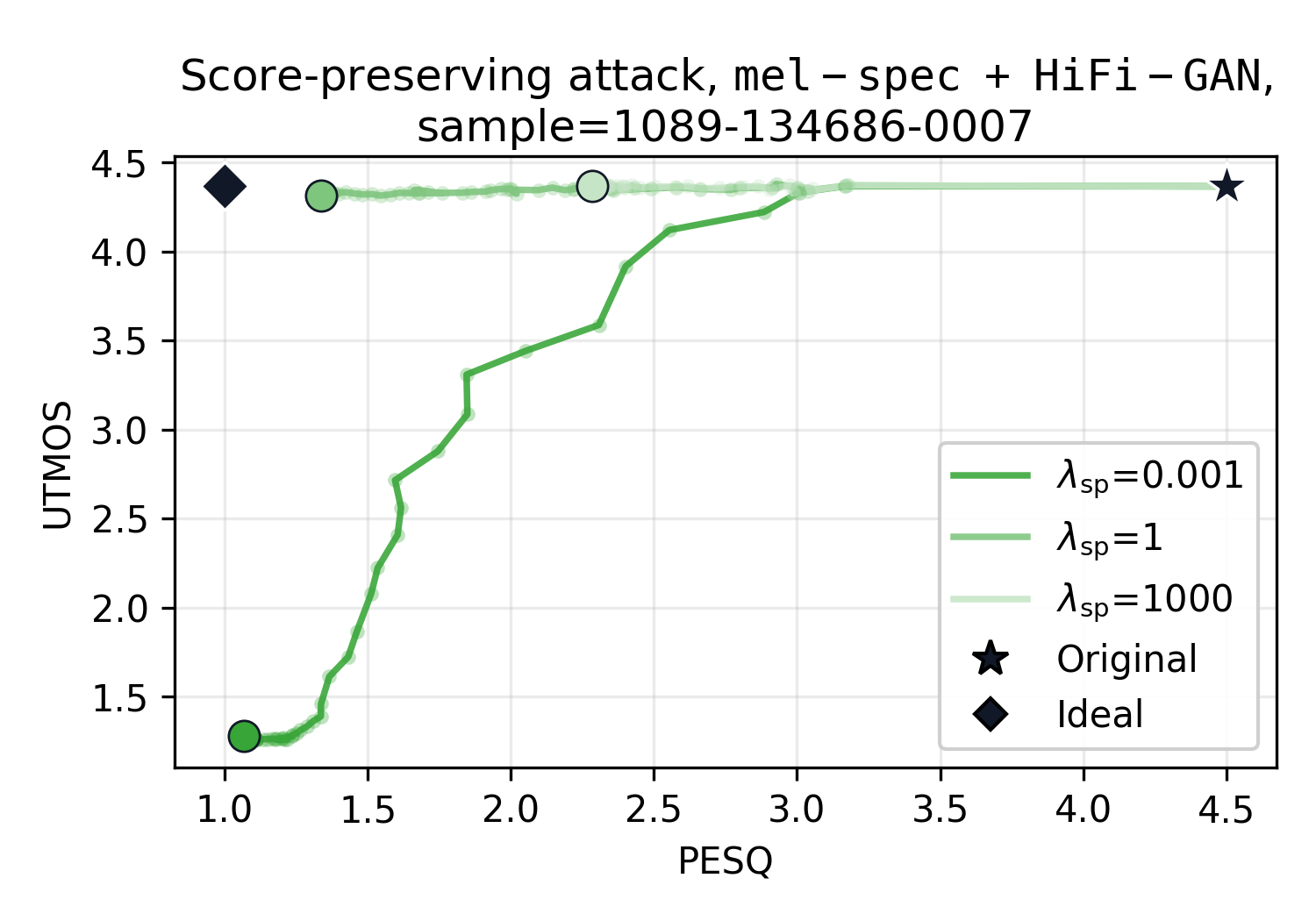}
    \caption{Optimization dynamics for the \mhfg space across different lambda weights.}
    \label{fig:score-preserve-lambda-compare}
  \end{subfigure}

  \vspace{0.75em}

  \begin{subfigure}{\linewidth}
    \centering
    \includegraphics[width=0.85\linewidth]{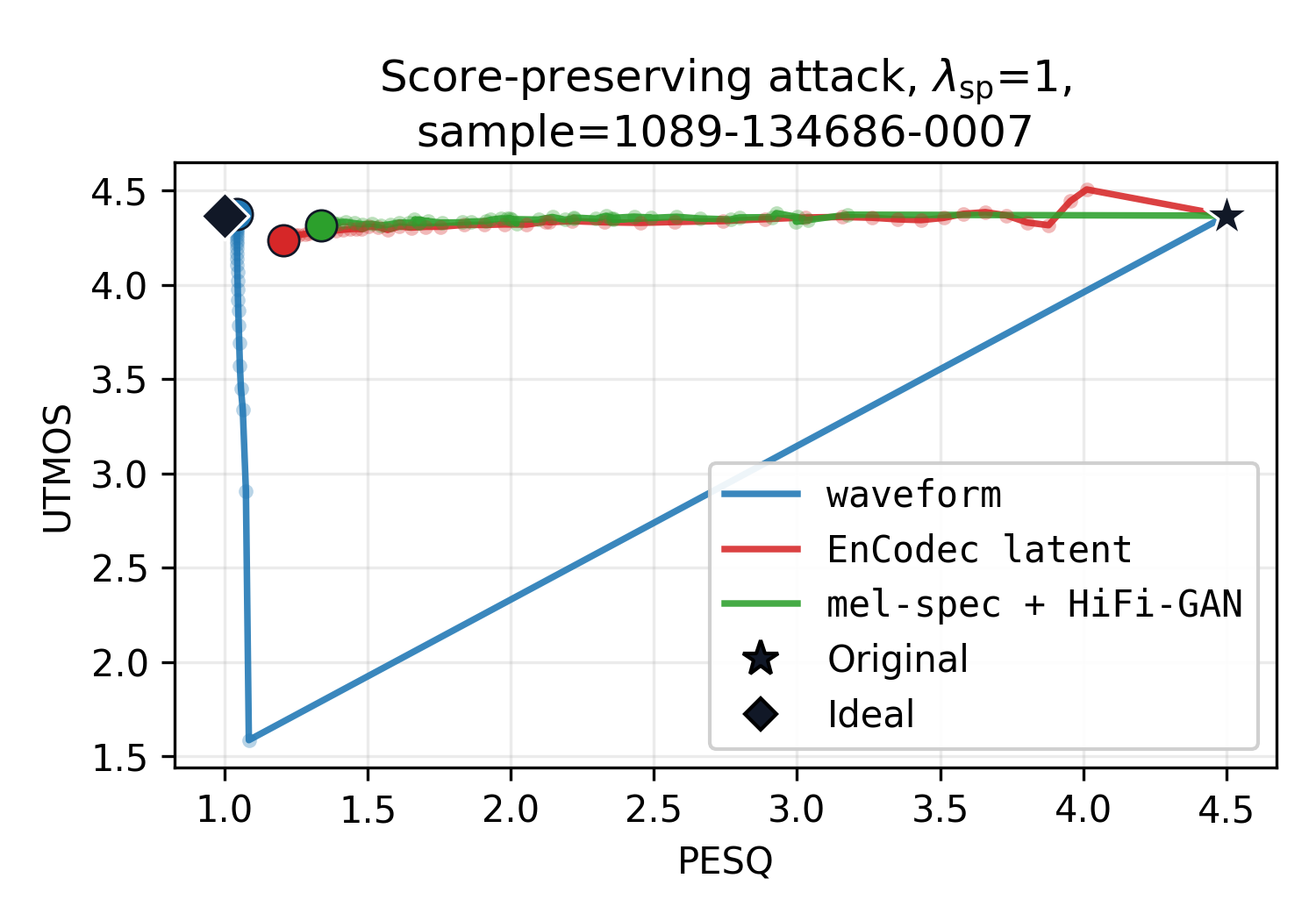}
    \caption{Optimization dynamics across the \wav, \mhfg, and \encodec spaces at a fixed lambda weight of $1$.}
    \label{fig:score-preserve-space-compare}
  \end{subfigure}

  \caption{Optimization dynamics in a single optimization run in score-preserving attack. One single, identical LibriSpeech sample is used across the experiments.}
  \label{fig:score-preserve-optimization-dynamics}
\end{figure}

\subsubsection{Dynamics in a single optimization run}

We further plot the optimization dynamics inside a single score-preserving attack run in Fig.~\ref{fig:score-preserve-optimization-dynamics}. First, Fig.~\ref{fig:score-preserve-lambda-compare} compares the dynamics with different $\lsp$ values in the \mhfg space. With a properly large $\lsp$, the score-preserving penalty was strong enough to ``support'' the optimization to reach the desired score-preserving region. In constrast, a small $\lsp$ makes the optimization degenerate to pure PESQ minimization, leading to a behavior similar to that described in Sec.~\ref{ssec:noise-baseline}.

Figure~\ref{fig:score-preserve-space-compare} compares the dynamics of all three attack spaces with $\lsp=1$. For the \mhfg and \encodec spaces, we observe a ``smooth'' horizontal movement, showing a gradual decrease in PESQ while maintaining the UTMOS score. What was interesting was the dynamics of the \wav space: at the very first step, the sample became severely ``contaminated'', reaching extremely low UTMOS and PESQ scores. Upon manual inspection, the audio sample at this point contained strong white noise. However, the optimization was able to ``micro-adjust'' the noise and ``climb'' up to the desired low-PESQ, high-UTMOS region.

Because HiFi-GAN and EnCodec were trained to generate speech, their output spaces are implicitly biased toward a speech-like manifold. Perturbations in these spaces thus tend to change the signal through speech-structured distortions rather than arbitrary sample-level noise, making it easier to gradually reduce PESQ while keeping UTMOS stable. In contrast, optimization in the \wav space is unconstrained and can immediately ``leave'' the speech manifold, producing severe contamination before the optimizer finds a noisy region with high UTMOS score. This interpretation is only a hypothesis; experimentally examining this mechanism is beyond the scope of this work.

\begin{figure}
    \centering
    
    \includegraphics[width=\linewidth]{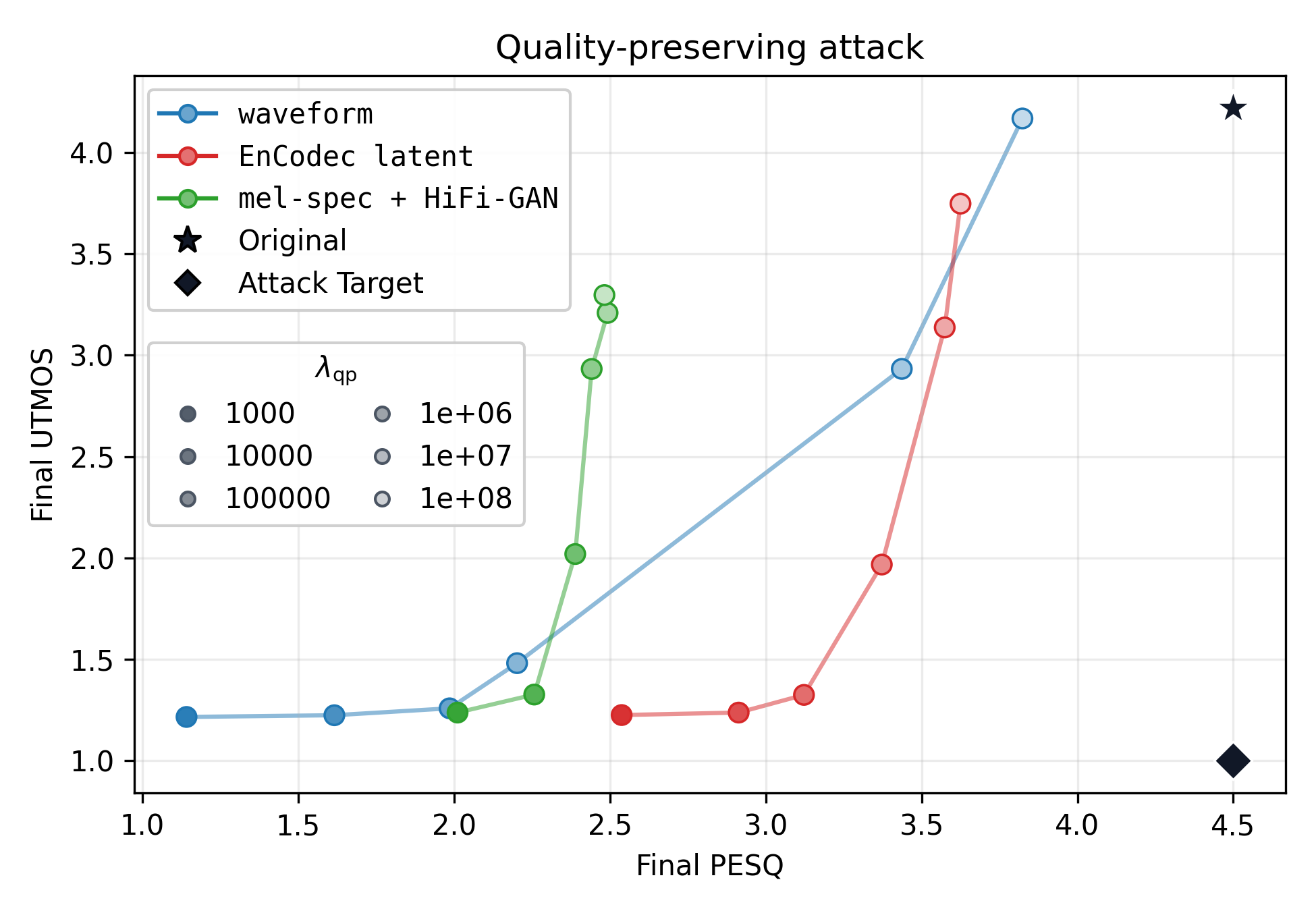}
    \caption{Quality-preserving attack results.}

    \label{fig:quality-preserving-results}
\end{figure}
\begin{figure*}[t]
    \centering
    \begin{subfigure}[t]{0.32\textwidth}
        \centering
        \includegraphics[width=\linewidth]{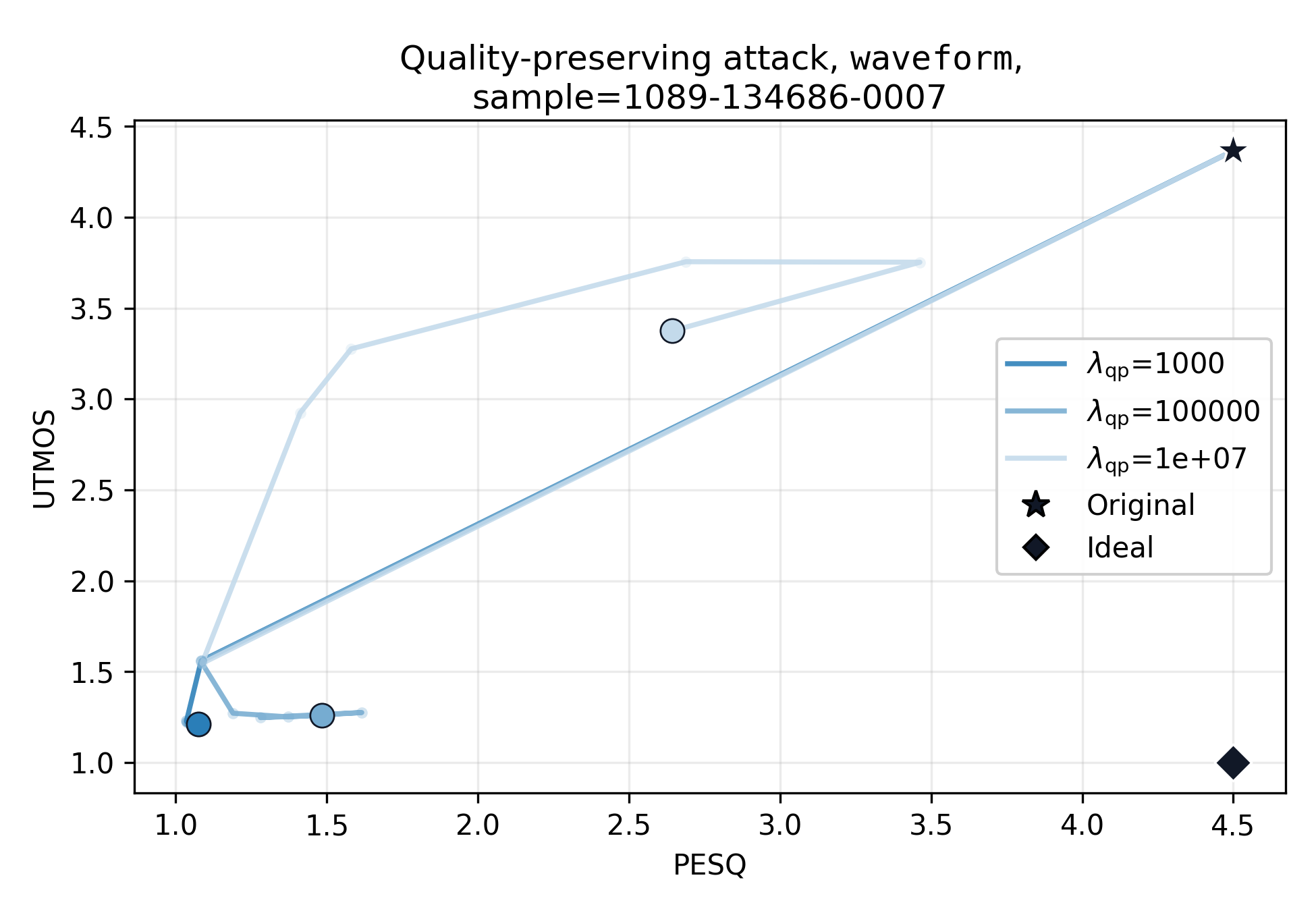}
        \caption{\wav}
        \label{fig:qp-dynamics-lambda-waveform}
    \end{subfigure}
    \hfill
    \begin{subfigure}[t]{0.32\textwidth}
        \centering
        \includegraphics[width=\linewidth]{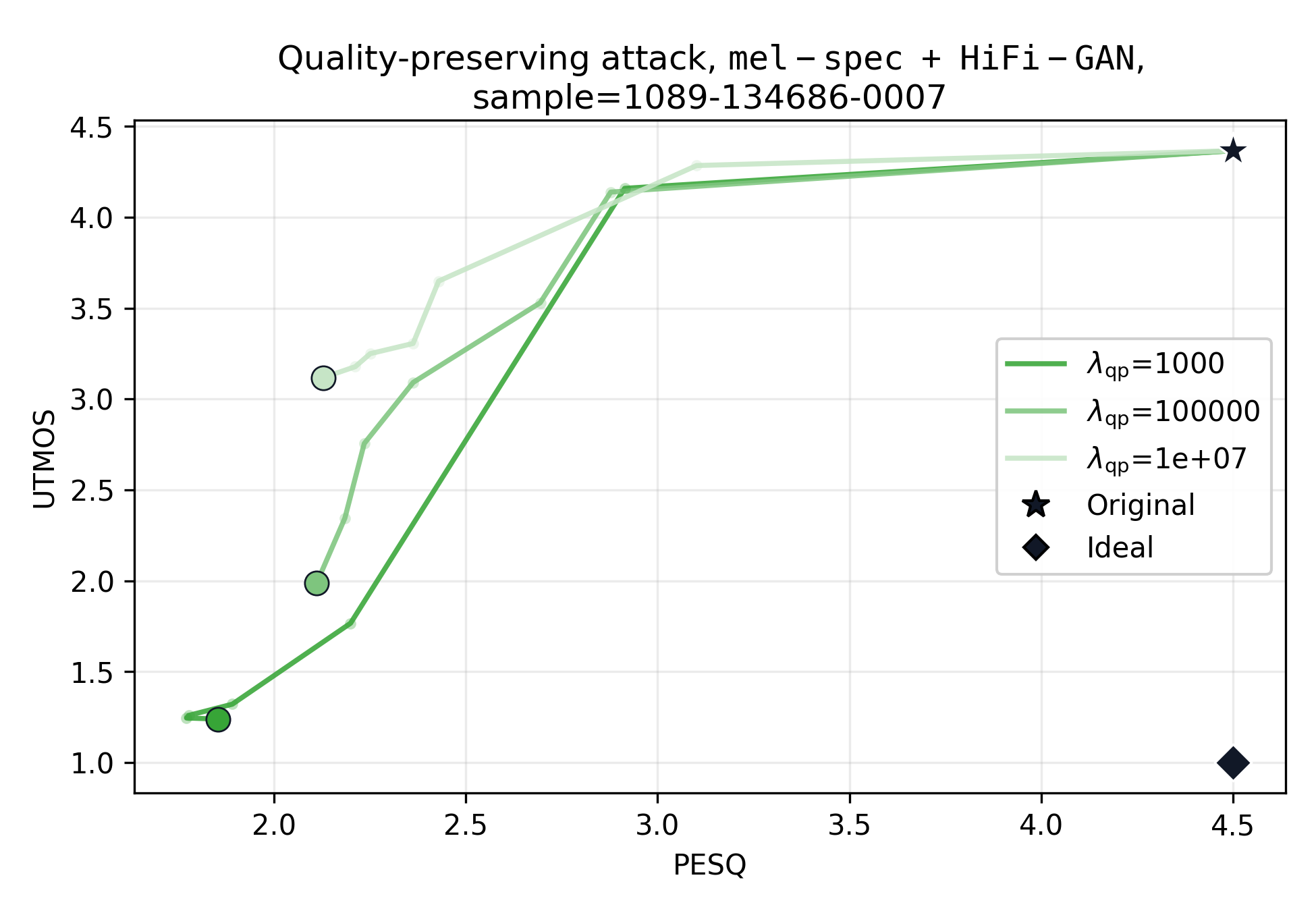}
        \caption{\mhfg}
        \label{fig:qp-dynamics-lambda-hifigan}
    \end{subfigure}
    \hfill
    \begin{subfigure}[t]{0.32\textwidth}
        \centering
        \includegraphics[width=\linewidth]{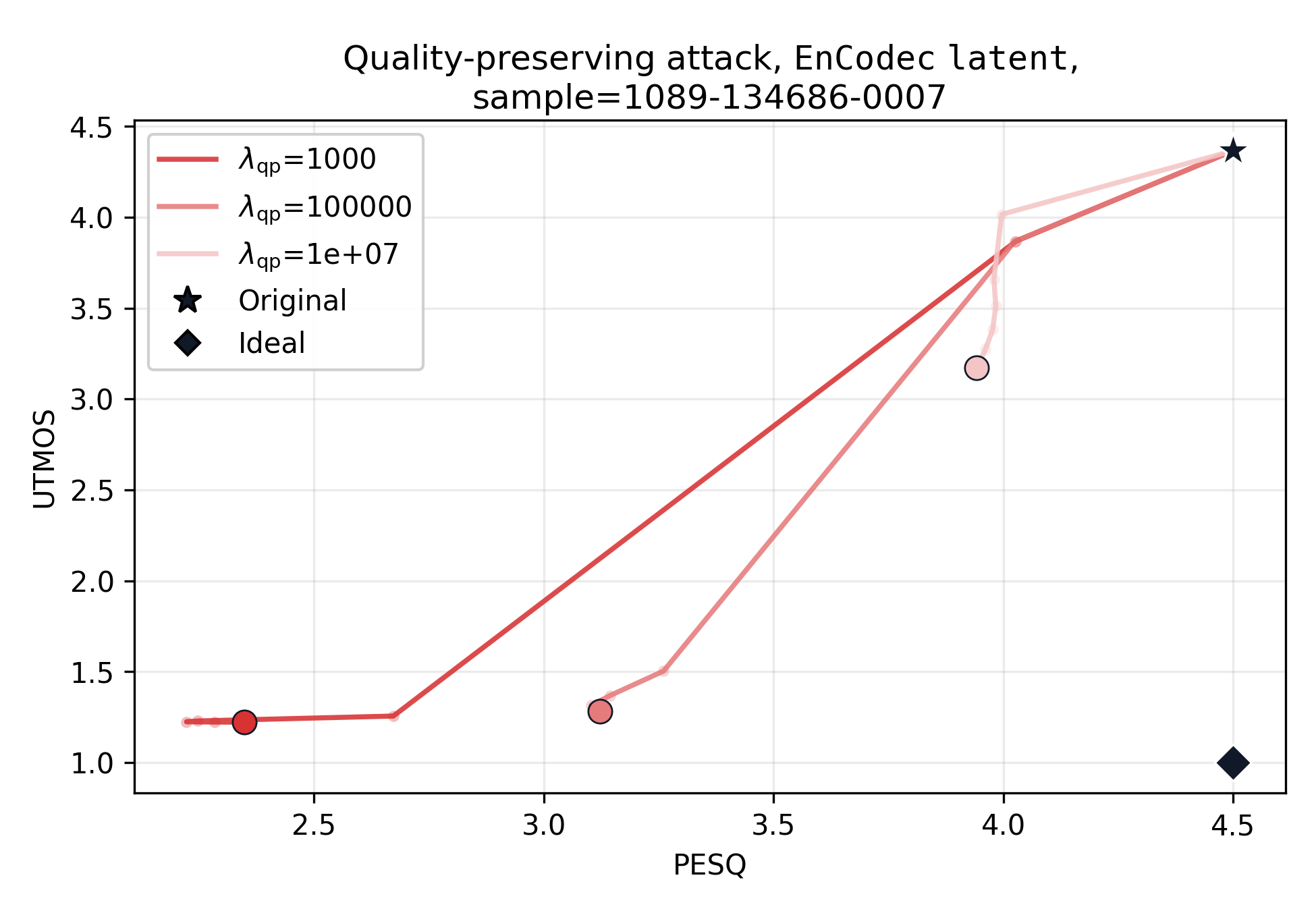}
        \caption{\encodec}
        \label{fig:qp-dynamics-lambda-encodec}
    \end{subfigure}
    \caption{
        Effect of the quality-preserving penalty weight $\lqp$ on optimization dynamics within each attack space.
        Each panel fixes the attack space and compares trajectories produced by different values of $\lqp$ for the same utterance.
        Larger $\lqp$ changes the trade-off between reducing predicted quality and preserving perceptual similarity, resulting in distinct paths in the PESQ--UTMOS plane.
    }
    \label{fig:qp-dynamics-lambda-compare}
\end{figure*}

\subsection{Quality-preserving attack results}
\label{ssec:quality-preserving-attack-results}

\subsubsection{Results from different attack spaces}

We next turn to the second attack direction, the quality-preserving attack. Fig.~\ref{fig:quality-preserving-results} shows the results obtained from different attack spaces and different values of $\lqp$, which controls the strength of the quality-preservation penalty in Eq.~\ref{eq:quality-preserving-attack}. In contrast to the score-preserving attack results in Sec.~\ref{ssec:score-preserving-attack-results}, where samples approached the target region, generating samples close to the desired quality-preserving region was substantially more difficult.

To quantify the degree of attack success, we examine $\dqp=\text{PESQ}-\text{UTMOS}$. An ideal attack would achieve $\dqp=3.5$ with $\text{PESQ}=4.5$ and $\text{UTMOS}=1.0$, and a larger value indicates a more successful attack. The most successful setting for each attack space is summarized as follows:
\begin{itemize}
    \item \wav: $\lqp = 1 \times 10^6$, $\text{(PESQ, UTMOS)} = (2.20, 1.48)$, and $\dqp=0.72$.
    \item \mhfg: $\lqp = 1 \times 10^4$, $\text{(PESQ, UTMOS)} = (2.26, 1.33)$, and $\dqp=0.93$.
    \item \encodec: $\lqp = 1 \times 10^5$, $\text{(PESQ, UTMOS)} = (3.12, 1.32)$, and $\dqp=1.80$.
\end{itemize}
These results indicate that optimization in the \encodec latent space provides the most promising results among the three attack spaces. However, manual inspection showed that even the best attacked samples in each space were perceptually different from the original samples, suggesting that fully quality-preserving attacks remain difficult.

\subsubsection{Dynamics in a single optimization run}

Figure~\ref{fig:qp-dynamics-lambda-compare} shows the optimization dynamics inside a single quality-preserving attack run across the three spaces. First, looking at the \wav results in Fig.~\ref{fig:qp-dynamics-lambda-waveform}, we observed a phenomenon similar to that in Fig.~\ref{fig:score-preserve-space-compare}: the sample reached extremely low UTMOS and PESQ scores at the very first step, regardless of the $\lqp$ value. Then, if the $\lqp$ value (i.e., the quality-preserving penalty) is too small, the optimization lacks the necessary gradient strength to recover. However, even when the $\lqp$ value is large—and the optimization mechanics successfully drive the coordinate toward the correct direction by increasing the perceived quality (PESQ) -- the UTMOS score inevitably increases as well. Looking at Figs.~\ref{fig:qp-dynamics-lambda-hifigan} and~\ref{fig:qp-dynamics-lambda-encodec}, the \mhfg and \encodec spaces demonstrated a more similar behavior, where the $\lqp$ value controls the magnitude of the suppressed quality drop.

\begin{table}[t]
\centering
\caption{UTMOS, PESQ and perceived quality (with 95\% confidence interval) results obtained from the MOS test for the quality-preserving attack.}
\label{tab:listening-test}

\scriptsize

\begin{tabular}{lcccc}
\toprule
System & Parameter & UTMOS & PESQ & Perceived quality \\
\midrule
Original & -- & 4.30 & 4.50 & 4.46 $\pm$ 0.12 \\
Random noise & $\text{amp}=0.01$ & 2.51 & 1.32 & 2.01 $\pm$ 0.17 \\
\midrule
\multirow{2}{*}{\wav}
  & $\lqp = 1 \times 10^7$ & 3.12 & 3.58 & 3.61 $\pm$ 0.12 \\
  & $\lqp = 1 \times 10^6$ & 1.50 & 2.35 & 2.69 $\pm$ 0.12 \\
\multirow{2}{*}{\shortstack[l]{\texttt{mel-spec}\\\hspace{0.7em}\texttt{+ HiFi-GAN}}}
  & $\lqp = 1 \times 10^5$ & 2.16 & 2.34 & 2.24 $\pm$ 0.14 \\
  & $\lqp = 1 \times 10^4$ & 1.35 & 2.19 & 1.96 $\pm$ 0.13 \\
\multirow{2}{*}{\shortstack[l]{\texttt{EnCodec}\\\hspace{0.7em}\texttt{latent}}}
  & $\lqp = 1 \times 10^6$ & 2.01 & 3.41 & 3.50 $\pm$ 0.16 \\
  & $\lqp = 1 \times 10^5$ & 1.36 & 3.21 & 3.11 $\pm$ 0.14 \\
\bottomrule
\end{tabular}


\end{table}

\subsection{Listening test results}

For the listening test, from the results in Sec.~\ref{ssec:score-preserving-attack-results}, we found that score-preserving samples exhibited severe degradation under PESQ and manual inspection. Thus, we focus the listening test on the more ambiguous quality-preserving setting.
We conducted a five-point scale MOS test to assess the perceived quality of ten randomly picked samples from each of the following systems: the original sample, the random noise perturbation baseline, and quality-preseving attacked $\wav$, $\mhfg$ and $\encodec$, each with two different $\lqp$ values. The detailed setting can be found in Tab.~\ref{tab:listening-test}. We recuited fourteen participants, where every single participant listened to each of the 80 samples.

Table~\ref{tab:listening-test} shows the results. We found that the perceived quality scores had a consistent trend with PESQ, with both sample-level and system-level linear correlation coefficients at 0.90 and 0.96, respectively. This result justifies the usage of PESQ as a proxy perceived quality metric. We also confirmed, via two-sided Wilcoxon signed-rank tests, that the attacked samples are statistically different from the original samples with $p$-values all smaller than $1\times 10^{-14}$, suggesting that fully quality-preserving attacks were not achievable against UTMOS.





\section{Conclusions and future work}


In this paper, we investigated the robustness of UTMOS, a widely used DNN-based speech quality assessment model, through input optimization on three spaces: \wav, \mhfg and \encodec.
Experimental results showed that UTMOS is clearly vulnerable to score-preserving attacks: severely degraded samples can still receive high predicted quality scores. In contrast, quality-preserving attacks were substantially more difficult, and the listening test confirmed that the attacked samples remained perceptually different from the original samples. Among the three attack spaces, optimization in the \encodec space provided the most promising results for the quality-preserving direction.
These results suggest that UTMOS may be particularly unreliable when used as an optimization target or reward, because high predicted scores can be maintained even under severe perceptual degradation. This does not invalidate its use as a passive evaluation metric, but it indicates that UTMOS scores should be interpreted with caution, especially in settings where systems are directly or indirectly optimized against the metric. More broadly, our findings suggest that robustness issues in DNN-based SQA models are not limited to naturally occurring out-of-domain samples, but can also be exposed through adversarially constructed examples.

As future work, we are first interested in extending the analysis beyond UTMOS. In particular, we plan to examine models trained on noisy and enhanced speech \cite{nisqa, dnsmos}, as well as models trained on broader collections of speech quality data \cite{mos-bench, urgentmos}. We are also interested in attacks that start from low-quality examples. Such attacks can be interpreted as a form of ``speech enhancement'' guided by SQA models, where the goal is to improve the predicted quality score through input optimization. Finally, we are interested in attacking non-score-based SQA models. For example, there is an increasing trend toward evaluating spoken dialogue systems using SQA models that output natural language descriptions, many of which are based on LLMs \cite{alld, qualispeech, speechllm-as-judges}. This is in spirit similar to works on attacking LLM-as-a-judge \cite{attack-llm-as-a-judge}, and we plan to explore methods such as prompt injection attacks \cite{prompt-injection-attack}.

\ifCLASSOPTIONpeerreview
\else

\section*{Acknowledgment}

This work was partly supported by JSPS KAKENHI Grant Number 25K00143, Japan, and in part by the BRIDGE Program (R7-H05), implemented by the Cabinet Office, Government of Japan.
We would like to thank Dr. Erica Cooper from NICT, Japan and Mr. Atsushi Miyashita from Nagoya University, Japan for the fruitful discussion.
The authors used ChatGPT 5.5 and Codex to (1) help generate source codes for the experiments (2) polish the English grammar and word usage of the manuscript. The idea is original, and authors take full responsibility for the contents in the manuscript.

\fi
\bibliographystyle{IEEEtran}
\bibliography{ref}

\end{document}